\documentclass[conference, 10pt]{IEEEtran}

\usepackage[cmex10]{amsmath}  
\usepackage{amssymb}
\usepackage{graphicx,color}   

\usepackage{ifxetex}
\ifxetex
\usepackage{xunicode}
\else
\usepackage[utf8]{inputenc}
\usepackage[T1]{fontenc}
\usepackage{microtype}
\fi

\newtheorem{theorem}{Theorem}
\newtheorem{remark}{Remark}
\newtheorem{lemma}{Lemma}
\newtheorem{definition}{Definition}

\newcommand{\mbf}[1]{\ensuremath{\boldsymbol{#1}}}

\def\gap{.93ex}
  \abovedisplayskip\gap
  \belowdisplayskip\gap
  \abovedisplayshortskip\gap
  \belowdisplayshortskip\gap

\begin{document}

\title{The Finite Field Multi-Way Relay Channel with Correlated Sources: Beyond Three Users}
\author{\IEEEauthorblockN{Lawrence Ong$^\dag$, Roy Timo$^\ddag$, Sarah J.\ Johnson$^\dag$}
\IEEEauthorblockA{$^\dag$School of Electrical Engineering and Computer Science, 
The University of Newcastle, Australia\\
$^\ddag$Institute for Telecommunications Research, University of South Australia, Australia\\
Email: lawrence.ong@cantab.net, roy.timo@unisa.edu.au, sarah.johnson@newcastle.edu.au}
\and
\IEEEauthorblockN{}
\IEEEauthorblockA{}
}
\maketitle

\begin{abstract}
The multi-way relay channel (MWRC) models cooperative communication networks in which many users exchange messages via a relay. In this paper, we consider the finite field MWRC with correlated messages. The problem is to find all achievable rates, defined as the number of channel uses required per reliable exchange of message tuple. For the case of three users, we have previously established that for a special class of source distributions, the set of all achievable rates can be found [Ong et al., ISIT 2010]. The class is specified by an almost balanced conditional mutual information (ABCMI) condition. In this paper, we first generalize the ABCMI condition to the case of more than three users. We then show that if the sources satisfy the ABCMI condition, then the set of all achievable rates is found and can be attained using a separate source-channel coding architecture. 

\end{abstract}

\section{Introduction}

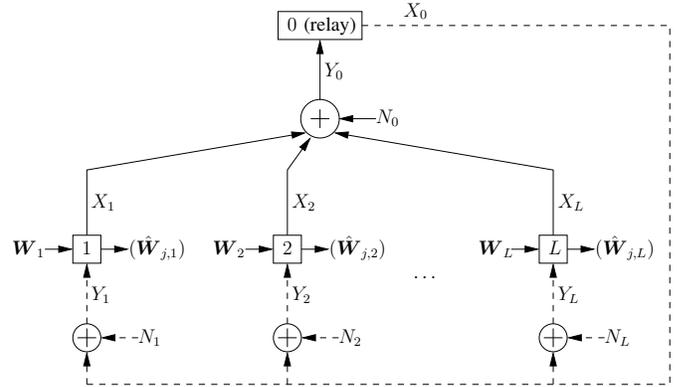
\begin{figure}[t]
\centering
\resizebox{\linewidth}{!}{ 
\begin{picture}(0,0)%
\includegraphics{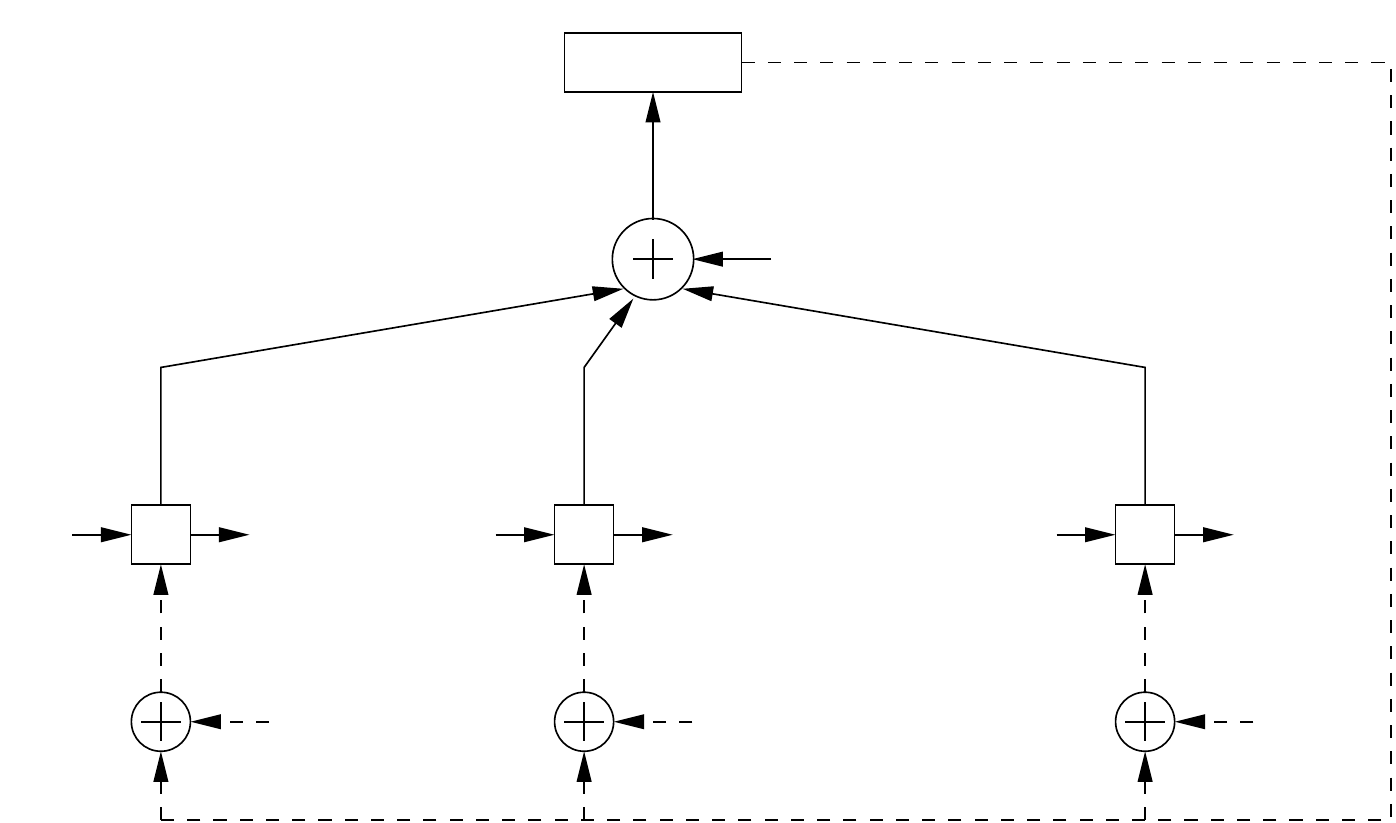}%
\end{picture}%
\setlength{\unitlength}{4144sp}%
\begingroup\makeatletter\ifx\SetFigFont\undefined%
\gdef\SetFigFont#1#2#3#4#5{%
  \fontsize{#1}{#2pt}%
  \fontfamily{#3}\fontseries{#4}\fontshape{#5}%
  \selectfont}%
\fi\endgroup%
\begin{picture}(6372,3750)(751,-3223)
\put(6481,-2806){\makebox(0,0)[lb]{\smash{{\SetFigFont{12}{14.4}{\familydefault}{\mddefault}{\updefault}{\color[rgb]{0,0,0}$N_L$}%
}}}}
\put(3421,209){\makebox(0,0)[lb]{\smash{{\SetFigFont{12}{14.4}{\familydefault}{\mddefault}{\updefault}{\color[rgb]{0,0,0}$0$  (relay)}%
}}}}
\put(1441,-1951){\makebox(0,0)[lb]{\smash{{\SetFigFont{12}{14.4}{\familydefault}{\mddefault}{\updefault}{\color[rgb]{0,0,0}$1$}%
}}}}
\put(766,-1951){\makebox(0,0)[lb]{\smash{{\SetFigFont{12}{14.4}{\familydefault}{\mddefault}{\updefault}{\color[rgb]{0,0,0}$\mbf{W}_1$}%
}}}}
\put(3376,-1951){\makebox(0,0)[lb]{\smash{{\SetFigFont{12}{14.4}{\familydefault}{\mddefault}{\updefault}{\color[rgb]{0,0,0}$2$}%
}}}}
\put(2701,-1951){\makebox(0,0)[lb]{\smash{{\SetFigFont{12}{14.4}{\familydefault}{\mddefault}{\updefault}{\color[rgb]{0,0,0}$\mbf{W}_2$}%
}}}}
\put(5941,-1951){\makebox(0,0)[lb]{\smash{{\SetFigFont{12}{14.4}{\familydefault}{\mddefault}{\updefault}{\color[rgb]{0,0,0}$L$}%
}}}}
\put(5266,-1951){\makebox(0,0)[lb]{\smash{{\SetFigFont{12}{14.4}{\familydefault}{\mddefault}{\updefault}{\color[rgb]{0,0,0}$\mbf{W}_L$}%
}}}}
\put(3826,-1951){\makebox(0,0)[lb]{\smash{{\SetFigFont{12}{14.4}{\familydefault}{\mddefault}{\updefault}{\color[rgb]{0,0,0}$(\hat{\mbf{W}}_{j,2})$}%
}}}}
\put(6391,-1951){\makebox(0,0)[lb]{\smash{{\SetFigFont{12}{14.4}{\familydefault}{\mddefault}{\updefault}{\color[rgb]{0,0,0}$(\hat{\mbf{W}}_{j,L})$}%
}}}}
\put(1891,-1951){\makebox(0,0)[lb]{\smash{{\SetFigFont{12}{14.4}{\familydefault}{\mddefault}{\updefault}{\color[rgb]{0,0,0}$(\hat{\mbf{W}}_{j,1})$}%
}}}}
\put(4636,-2221){\makebox(0,0)[lb]{\smash{{\SetFigFont{12}{14.4}{\familydefault}{\mddefault}{\updefault}{\color[rgb]{0,0,0}$\dotsm$}%
}}}}
\put(4546,344){\makebox(0,0)[lb]{\smash{{\SetFigFont{12}{14.4}{\familydefault}{\mddefault}{\updefault}{\color[rgb]{0,0,0}$X_0$}%
}}}}
\put(3781,-241){\makebox(0,0)[lb]{\smash{{\SetFigFont{12}{14.4}{\familydefault}{\mddefault}{\updefault}{\color[rgb]{0,0,0}$Y_0$}%
}}}}
\put(4276,-691){\makebox(0,0)[lb]{\smash{{\SetFigFont{12}{14.4}{\familydefault}{\mddefault}{\updefault}{\color[rgb]{0,0,0}$N_0$}%
}}}}
\put(3466,-1501){\makebox(0,0)[lb]{\smash{{\SetFigFont{12}{14.4}{\familydefault}{\mddefault}{\updefault}{\color[rgb]{0,0,0}$X_2$}%
}}}}
\put(1531,-1501){\makebox(0,0)[lb]{\smash{{\SetFigFont{12}{14.4}{\familydefault}{\mddefault}{\updefault}{\color[rgb]{0,0,0}$X_1$}%
}}}}
\put(6031,-1501){\makebox(0,0)[lb]{\smash{{\SetFigFont{12}{14.4}{\familydefault}{\mddefault}{\updefault}{\color[rgb]{0,0,0}$X_L$}%
}}}}
\put(6031,-2401){\makebox(0,0)[lb]{\smash{{\SetFigFont{12}{14.4}{\familydefault}{\mddefault}{\updefault}{\color[rgb]{0,0,0}$Y_L$}%
}}}}
\put(3466,-2401){\makebox(0,0)[lb]{\smash{{\SetFigFont{12}{14.4}{\familydefault}{\mddefault}{\updefault}{\color[rgb]{0,0,0}$Y_2$}%
}}}}
\put(1531,-2401){\makebox(0,0)[lb]{\smash{{\SetFigFont{12}{14.4}{\familydefault}{\mddefault}{\updefault}{\color[rgb]{0,0,0}$Y_1$}%
}}}}
\put(1981,-2806){\makebox(0,0)[lb]{\smash{{\SetFigFont{12}{14.4}{\familydefault}{\mddefault}{\updefault}{\color[rgb]{0,0,0}$N_1$}%
}}}}
\put(3916,-2806){\makebox(0,0)[lb]{\smash{{\SetFigFont{12}{14.4}{\familydefault}{\mddefault}{\updefault}{\color[rgb]{0,0,0}$N_2$}%
}}}}
\end{picture}%
}
\caption{The finite field MWRC in which $L$ users (nodes $1,2,\dotsc, L$) exchange correlated messages through a relay (node $0$). The uplink channel is marked with solid lines and the downlink channel with dotted lines.} 
\label{fig:mwrc}
\end{figure}

This paper investigates multi-way relay channels (MWRCs) where multiple users exchange correlated data via a relay. More specifically, each user is to send its data to all other users. There is no direct link among the users, and hence the users first transmit to a relay, which processes its received information and transmits back to the users (refer to Fig.~\ref{fig:mwrc}). The purpose of this paper is to find the set of all achievable rates, which are defined as the number of channel uses required to reliably (in the usual Shannon sense) exchange each message tuple.

The joint source-channel coding problem in Fig.~\ref{fig:mwrc} includes, as special cases, the source coding work of Wyner et al.~\cite{wynerwolf02} and the channel capacity work of Ong et al.~\cite{ongmjohnsonit11}. Separately, the Shannon limits of source coding~\cite{wynerwolf02} (through noiseless channel) and of channel coding~\cite{ongmjohnsonit11} (with independent sources) are well established. However, these limits have not yet been discovered for noisy channels with correlated sources in general. For three users, Ong et al.~\cite{ongtimo11isit} gave sufficient conditions for reliable communication using the separate source-channel coding paradigm. The key result of~\cite{ongtimo11isit} was to show that these sufficient conditions are also necessary for a special class of  source distributions, hence giving the set of all achievable rates. The class was characterized by an almost balanced conditional mutual information (ABCMI) condition. This paper extends the ABCMI concept to more than three users, and shows that if the sources have ABCMI, then the set of all achievable rates can be found.
While the ABCMI condition for the three-user case is expressed in terms of the standard Shannon information measure, we will use {\em I}-Measure~\cite[Ch.\ 3]{yeung08} for more then three users---using Shannon's measure is possible but the expressions would be much more complicated. 

Though the ABCMI condition limits the class of sources for which the set of all achievable rates is found, this paper provides the following insights:
(i) As achievability is derived based on a separate source-channel coding architecture, we show that source-channel separation is optimal for finite field MWRC with sources having ABCMI. (ii) Since ABCMI are constraints on the sources, the results in this paper are potentially useful for other channel models (not restricted to the finite field model). (iii) This paper highlights the usefulness of the {\em I}-Measure as a complement to the standard Shannon information measure.

\section{Main Results}

\subsection{The Network Model}

\subsubsection{Sources} \label{sec:sources}
Consider $m$ independent and identically distributed (i.i.d.) drawings of a tuple of correlated discrete finite random variables $(W_1, W_2, \dotsc ,W_L)$, i.e., $\{(W_1[t],W_2[t],\dotsc,W_L[t])\}_{t=1}^m$.
The message of user~$i$ is given by
$\boldsymbol{W}_i = (W_i[1], W_i[2],\dotsc,W_i[m])$.

\subsubsection{Channel}
Each channel use of the finite field MWRC consists of an uplink and $L$ downlinks, characterized by
\begin{align}
\text{Uplink:} \quad &Y_0 = X_1 \oplus X_2 \oplus \dotsm \oplus X_L \oplus N_L  \label{eq:uplink} \\
\text{Downlinks:}  \quad &Y_i = X_0 \oplus N_i, \quad \text{for all }i \in \{1,2,\dotsc,L\}, \label{eq:downlink}
\end{align}
where $X_\ell,Y_\ell,N_\ell$, for all $\ell \in \{0,1,\dotsc, L\}$, each take values in a finite field $\mathcal{F}$, $\oplus$ denotes addition over $\mathcal{F}$, $X_\ell$ is the channel input from node $\ell$, $Y_\ell$ is the channel output received by node $\ell$, and $N_\ell$ is the receiver noise at node $\ell$. The noise $N_\ell$ is arbitrarily distributed, but is i.i.d. for each channel use. We have used the subscript $i$ to denote a user and the subscript $\ell$ to denote a node (which can be a user or the relay).

\subsubsection{Block Codes (joint source-channel codes with feedback)}
Consider block codes for which the users exchange $m$ message tuples in $n$ channel uses. [Encoding:] The $t$-th transmitted channel symbol of node $\ell$ is a function of its message and the $(t-1)$ symbols it previously observed on the downlink: $X_\ell[t] = f_\ell[t] (\boldsymbol{W}_\ell, Y_\ell[1], Y_\ell[2], \dotsc, Y_\ell[t-1])$, for all $\ell \in \{0,1,\dotsc, L\}$ and for all $t \in \{1,2,\dotsc,n\}$. As the relay has no message, we set $\boldsymbol{W}_0 = \varnothing$. [Decoding:]
The messages decoded by user $i$ are a function of its message and the symbols it observed on the downlink: $(\hat{\boldsymbol{W}}_{j,i}: \forall j \in \{1,\dotsc,L\} \setminus \{i\}) = h_i (\boldsymbol{W}_i, Y_i[1],Y_i[2],\dotsc,Y_i[n])$, for all $i \in \{1,2,\dotsc, L\}$.

\subsubsection{Achievable Rate}
Let $P_\text{e}$ denote the probability that $\hat{\boldsymbol{W}}_{j,i} \neq \boldsymbol{W}_i$ for any $i \neq j$. The \emph{rate} (or bandwidth expansion factor) of the code is the ratio of channel symbols to source symbols, $\kappa = n/m$. The rate $\kappa$ is said to be {\em achievable} if the following is true: for any $\epsilon > 0$, there exists a block code, with $n$ and $m$ sufficiently large and $n/m = \kappa$, such that  $P_\text{e} < \epsilon$.

\subsection{Statement of Main Results}

\begin{theorem} \label{theorem:main}
Consider an $L$-user finite field MWRC with correlated sources. If the sources $(W_1,W_2,\ldots,W_L)$ have almost balanced conditional mutual information (ABCMI), then $\kappa$ is achievable if and only if
\begin{equation}
\kappa \geq \max_{i \in \{1,2,\dotsc,L\}} \frac{H(W_{\{1,2,\dotsc,L\}\setminus \{i\}}|W_i) }{ \log_2|\mathcal{F}| - \max\{ H(N_0), H(N_i)\}}. \label{eq:main}
\end{equation}
\end{theorem}

The ABCMI condition used in Theorem~\ref{theorem:main}  is rather technical and best defined using the $I$-measure~\cite[Ch.\ 3]{yeung08}. For this reason, we specify this condition later in Section~\ref{section:abcmi} after giving a brief review of the $I$-measure in Section~\ref{section:i-measure}. For now, it suffices to note that it is a non-trivial constraint placed on the joint distribution of $(W_1,W_2,\ldots,W_L)$.

The achievability (\emph{if} assertion) of Theorem~\ref{theorem:main} is proved using a separate source-channel coding architecture, which involves intersecting a certain Slepian-Wolf source-coding region with the finite field MWRC capacity region~\cite{ongmjohnsonit11}. The particular source-coding region of interest is the classic Slepian-Wolf region \cite{han80} with the total sum-rate constraint omitted; specifically, it is the set of all {\em source-coding rate} tuples $(r_1,r_2,\ldots,r_L)$ such that
\begin{equation}
\sum_{i \in \mathcal{S}} r_i \geq H( W_{\mathcal{S}} | W_{\{1,2,\dotsc,L\} \setminus \mathcal{S}} ) \label{eq:lemma}
\end{equation}
holds for all strict subsets $\mathcal{S} \subset \{1,2,\ldots,L\}$. The next theorem will be a critical step in the proof of Theorem~\ref{theorem:main}.
\begin{theorem} \label{theorem:tuple}
If $L$ arbitrarily correlated random variables $(W_1, W_2, \dotsc, W_L)$ have ABCMI, then we can find a non-negative real tuple $(r_1, r_2, \dotsc, r_L)$ such that
\begin{itemize}
\item[{[}$\mathsf{C1}${]}]  the inequality~\eqref{eq:lemma} holds for all subsets $\mathcal{S} \subset \{1,2,\dotsc, L\}$ for which $1 \leq |\mathcal{S}| \leq L-2$,
\item[{[}$\mathsf{C2}${]}] the inequality~\eqref{eq:lemma} holds {\em with equality} for all subsets $\mathcal{S} \subset \{1,2,\dotsc, L\}$ for which  $|\mathcal{S}| =L-1$.
\end{itemize}
\end{theorem}


\begin{remark}
Theorem~\ref{theorem:main} characterizes a class of sources (on any finite field MWRC) for which (i) the set of all achievable rates is known, and (ii) source-channel separation holds.
\end{remark}

\begin{remark}
Slepian-Wolf type constraints of the form~\eqref{eq:lemma} appear often in multi-terminal information theory. Since Theorem~\ref{theorem:tuple} applies directly to such constraints, it might be useful beyond its application here to the finite field MWRC.
\end{remark}

\begin{remark}
To prove Theorem~\ref{theorem:tuple}, we need to select $L$ non-negative numbers that  satisfy $(2^L-2)$ equations.
\end{remark}

\section{Definition of ABCMI}

\subsection{The {\itshape I}-Measure} \label{section:i-measure}

Consider $L$ jointly distributed random variables $(W_1, W_2,$ $\dotsc, W_L)$. The Shannon measures of these random variables can be efficiently characterized via set operations and the {\em I}-measure. For each random variable $W_i$, we define a (corresponding) set $\tilde{W}_i$. Let $\mathcal{F}_L$ be the field generated by $\{\tilde{W}_i\}$ using the usual set operations union $\cup$, intersection $\cap$, complement $^\text{c}$, and difference $-$.  The relationship between $\tilde{W}_i$ and $W_i$ is described by the {\itshape I}-Measure $\mu^*$ on $\mathcal{F}_L$, defined as~\cite[Ch.\ 3]{yeung08}
\begin{equation}
\mu^*(\tilde{W}_\mathcal{S}) = H(W_\mathcal{S}), \label{eq:i-measure}
\end{equation}
for any non-empty $\mathcal{S} \subseteq \{1,2,\dotsc, L\}$,
where $\tilde{W}_\mathcal{S} = \bigcup_{i \in \mathcal{S}} \tilde{W}_i$ and  $W_\mathcal{S} \triangleq \{W_i: i \in \mathcal{S}\}$. 

The {\em atoms} of $\mathcal{F}_L$ are sets of the form $\bigcap_{i=1}^L U_i$, where $U_i$ can either be $\tilde{W}_i$ or $\tilde{W}^\text{c}_i$. There are $2^L$ atoms in $\mathcal{F}_L$, and we denote the atoms by
\begin{equation}
a(\mathcal{K}) \triangleq \bigcap_{i \in \mathcal{K}} \tilde{W}_i - \bigcup_{j \in \mathcal{K}^\text{c}} \tilde{W}_j, \label{eq:atom}
\end{equation}
for all $\mathcal{K} \subseteq \{1,2,\dotsc,L\}$ where $\mathcal{K}^\text{c} \triangleq \{1,2,\dotsc,L\} \setminus \mathcal{K}$. Note that each atom corresponds to a unique $\mathcal{K} \subseteq \{1,2,\dotsc, L\}$. For the atom in \eqref{eq:atom}, we call $|\mathcal{K}|$ the {\em weight} of the atom. 


\begin{remark}
The {\em I}-measure of the atoms corresponds to the conditional mutual information of the variables. More specifically, $\mu^*(a(\mathcal{K}))$ is the mutual information among the variables $\{W_i: i \in \mathcal{K}\}$ conditioning on $\{W_j: j \in \mathcal{K}^\text{c}\}$. For example, if $L=4$, then $\mu^*(a(1,2)) = I(W_1;W_2|W_3,W_4)$, where $I(\cdot)$ is Shannon's measure of conditional mutual information.
\end{remark}

\subsection{Almost Balanced Conditional Mutual Information}  \label{section:abcmi}

For each $K \in \{1,2,\dotsc, L-1\}$, we define
\begin{align}
\overline{\mu}_K &\triangleq \max_{\substack{\mathcal{K} \subseteq \{1,2,\dotsc,L\} \\ \text{ s.t. } |\mathcal{K}| = K}} \mu^*( a(\mathcal{K}))\\
\underline{\mu}_K &\triangleq \min_{\substack{\mathcal{K} \subseteq \{1,2,\dotsc,L\} \\ \text{ s.t. } |\mathcal{K}| = K}}  \mu^*( a(\mathcal{K})),
\end{align}
i.e., atoms of weight $K$ with the largest and the smallest measures respectively.
With this, we define the ABCMI condition for $L$ random variables:
\begin{definition} \label{def:abcmi}
$(W_1,W_2,\dotsc,W_L)$ are said to have ABCMI if the following conditions hold:
\begin{equation*}
\overline{\mu}_{L-1} \leq \underline{\mu}_{L-1} \left(1 + \displaystyle \frac{1}{L-2} \right)
\end{equation*}
and
\begin{equation*}
\overline{\mu}_K \leq \underline{\mu}_K\left(1 + \displaystyle\frac{1}{K-1} \left[ {\displaystyle\min_{S \in \{K, K+1, \dotsc, L-2\}}\frac{\beta(L,S,K)}{\alpha(L,S,K) } }\right] \right),
\end{equation*}
for all $K \in \{2,3,\dotsc ,L-2\} $,
where
\begin{align}
\alpha(L,S,K) &\triangleq S \binom{L-1}{K} - (S-K)\binom{S}{K} \label{eq:alpha}\\
\beta(L,S,K) &\triangleq S \binom{L-1}{K} - (L-1) \binom {S}{K}, \label{eq:beta}
\end{align}
where $\binom{S}{K} \triangleq \frac{S!}{K!(S-K)!}$.

\end{definition}

\begin{remark}
The ABCMI condition requires that all atoms of the same weight (except for those with weight equal to zero, one, or $L$) have about the same {\itshape I}-measure.
\end{remark}
\begin{remark}
For any $L$, $S$, and $K$, such that $2 \leq K \leq S \leq L-2$, it can be  shown that $\alpha(L,S,K) \geq \beta(L,S,K) \geq 0$.
\end{remark}
\begin{remark}
For $L=3$, we have $\overline{\mu}_2  \leq 2 \underline{\mu}_2$, i.e.,   we recover the ABCMI condition for the three-user case~\cite{ongtimo11isit}.
\end{remark}


\section{Proof of Theorem~\ref{theorem:tuple}} \label{sec:1}



For the rest of this paper, we are interested in atoms only with  weight between one and $(L-1)$ inclusive. So, we define $\mathbb{K} \triangleq \{ \mathcal{K} \subset \{1,2,\dotsc, L\}: 1 \leq |\mathcal{K}| \leq L-1 \}$ and refer to $\mathbb{A} \triangleq \{ a(\mathcal{K}): \forall \mathcal{K} \in \mathbb{K}\}$ as the set of  all such atoms.

We propose to select $(r_1,r_2,\dotsc, r_L)$ in terms of the {\em I}-Measure:
\begin{equation}
r_i =
\sum_{\mathcal{K} \in \mathbb{K}} J_i (\mathcal{K}), \label{eq:rate-assignment}
\end{equation}
where
\begin{equation*}
J_i (\mathcal{K}) = \begin{cases}
+\frac{L- |\mathcal{K}|}{L-1} \mu^* ( a(\mathcal{K})), &\text{if } i \in \mathcal{K}\\
-\frac{|\mathcal{K}|-1}{L-1} \mu^* ( a(\mathcal{K})), &\text{otherwise, i.e., if } i \in \mathcal{K}^\text{c}.
\end{cases}
\end{equation*}

Each $r_i$ is chosen as the sum of the {\em weighted} (by a coefficient $+\frac{L- |\mathcal{K}|}{L-1}$ or $-\frac{|\mathcal{K}|-1}{L-1}$) {\itshape I}-measure of all atoms in $\mathbb{A}$. The assignments of $r_i$'s are depicted in Fig.~\ref{fig:atoms}. We term $J_i(\mathcal{K})$  the {\em contribution} from the atom $a(\mathcal{K})$ to $r_i$. Each contribution is represented by a {\em cell} in Fig.~\ref{fig:atoms}, and each $r_i$ by a column.

We now show that  conditions $\mathsf{C1}$ and $\mathsf{C2}$ in Theorem~\ref{theorem:tuple} hold when the $r_i$'s are chosen as per \eqref{eq:rate-assignment}.

\subsection{For $|\mathcal{S}| = L-1$:}

We first show that for any $i \in \{1,2,\dotsc,L\}$, we have 
\begin{equation}
\sum_{j \in \{1,2,\dotsc, L\} \setminus \{i\}} r_j  = H( W_{\{1,2,\dotsc, L\} \setminus \{i\}} | W_i). \label{eq:last-case}
\end{equation}

Since $r_i$'s are  defined in terms of {\em I}-Measure, we link the measure of atoms to the entropies of the corresponding random variables:
\begin{subequations}
\begin{align}
H(W_\mathcal{S} | W_{\mathcal{S}^\text{c}}) &= \mu^* ( \tilde{W}_\mathcal{S} - \tilde{W}_{\mathcal{S}^\text{c}}) \label{eq:measure}\\
&= \sum_{\text{non-empty }\mathcal{K} \subseteq \mathcal{S}} \mu^*( a(\mathcal{K})), \label{eq:set}
\end{align}
\end{subequations}
where \eqref{eq:measure} follows from \cite[eqn.~(3.43)]{yeung08}, and \eqref{eq:set} is obtained by counting all the atoms in the set $(\tilde{W}_\mathcal{S} - \tilde{W}_{\mathcal{S}^\text{c}})$.

\begin{figure}[t]
\centering
\includegraphics[width=7.7cm]{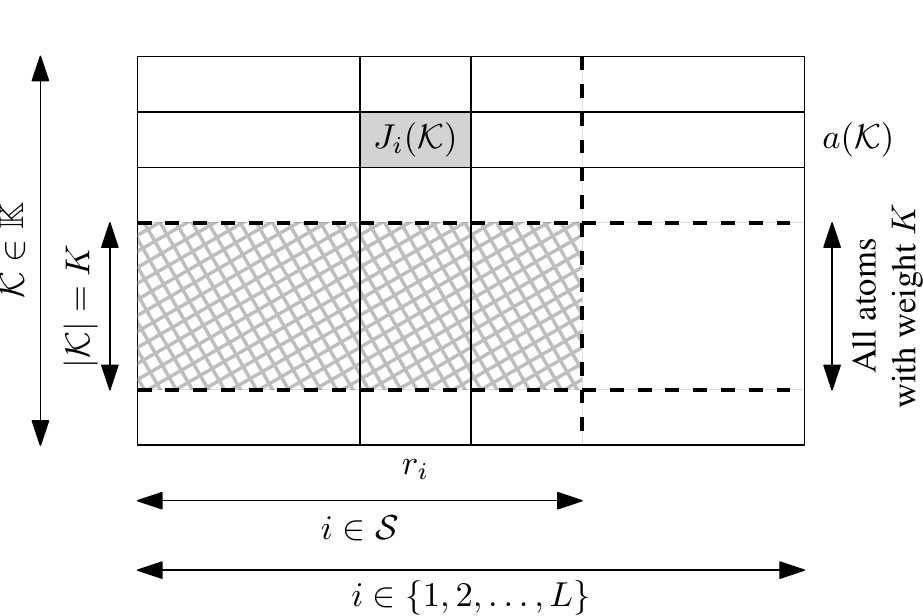} 
\caption{Each row represents the contributions from a unique atom, and $r_i$ is the summation of all cells, i.e., $\{J_i(\cdot)\}$, in the $i$-th column. The hashed region shows the contributions from all atoms with weight $K$ to all $r_i$'s for $i$ in some set $\mathcal{S}$.}
\label{fig:atoms}
\end{figure}

This means the RHS of \eqref{eq:last-case} equals $\sum_{\text{non-empty }\mathcal{K} \subseteq \{1,2,\dotsc, L\} \setminus \{i\}}\mu^*( a(\mathcal{K}))$. 

We now evaluate the LHS of \eqref{eq:last-case} for some fixed $i$. Consider some atom $a(\mathcal{K})$ where $i \notin \mathcal{K}$. We evaluate the contributions from this atom to $\boldsymbol{r}^{-i} \triangleq \{r_j: j \in \{1,2,\dotsc, L\} \setminus \{i\}\}$, i.e., one specific row in Fig.~\ref{fig:atoms} less the cell $J_i(\mathcal{K})$:
\begin{itemize}
\item $|\mathcal{K}|$ of the $(L-1)$ cells each contribute $\frac{L- |\mathcal{K}|}{L-1} \mu^* (a(\mathcal{K}))$.
\item The remaining $(L-1-|\mathcal{K}|)$ cells each contribute $-\frac{|\mathcal{K}|-1}{L-1} \mu^* (a(\mathcal{K}))$.
\end{itemize}
So, summing the contributions from $a(\mathcal{K})$ to $\boldsymbol{r}^{-i}$, we have
\begin{subequations}
\begin{align}
&\sum_{j \in \{1,2,\dotsc,L\} \setminus \{i\}} J_j(\mathcal{K}) \nonumber \\&= \left[ |\mathcal{K}|\frac{L- |\mathcal{K}|}{L-1}- (L-1-|\mathcal{K}|) \frac{|\mathcal{K}|-1}{L-1} \right]\mu^* (a(\mathcal{K}))\\
& = \mu^*(a(\mathcal{K})).
\end{align}
\end{subequations}

Consider some atom $a(\mathcal{K}')$ where $i \in \mathcal{K}'$. The contributions from this atom to $\boldsymbol{r}^{-i}$ are as follows (again, one specific row in Fig.~\ref{fig:atoms} less the cell $J_i(\mathcal{K}')$):
\begin{itemize}
\item $(|\mathcal{K}'|-1)$ of the $(L-1)$ cells each contribute $\frac{L- |\mathcal{K}'|}{L-1} \mu^* (a(\mathcal{K}'))$.
\item The remaining $(L-|\mathcal{K}'|)$ cells each contribute $-\frac{|\mathcal{K}'|-1}{L-1} \mu^* (a(\mathcal{K}'))$.
\end{itemize}
So, summing the contributions from $a(\mathcal{K}')$ to $\boldsymbol{r}^{-i}$, we have
\begin{align*}
&\sum_{j \in \{1,2,\dotsc,L\} \setminus \{i\}} J_j(\mathcal{K}') \nonumber \\ &= \left[ (|\mathcal{K}'|-1)\frac{L- |\mathcal{K}'|}{L-1} - (L-|\mathcal{K}'|) \frac{|\mathcal{K}'|-1}{L-1} \right] \mu^* (a(\mathcal{K}')) = 0.
\end{align*}

Combining the above results, we have, for a fixed $i$,
\begin{subequations}
\begin{align}
&\sum_{j \in \{1,2,\dotsc, L\} \setminus \{i\}} r_j \nonumber\\  &= \sum_{j \in \{1,2,\dotsc, L\} \setminus \{i\}}  \left[ \sum_{\substack{\mathcal{K} \in \mathbb{K} \\ \text{s.t. } i \notin \mathcal{K}}}  J_j(\mathcal{K})  + \sum_{\substack{\mathcal{K}' \in \mathbb{K}\\ \text{s.t. } i \in \mathcal{K}'}}  J_j(\mathcal{K}') \right] \\
& = \sum_{\substack{\mathcal{K} \in \mathbb{K}\\ \text{s.t. } i \notin \mathcal{K}}} \mu^*( a(\mathcal{K})) \\
& = \sum_{\text{non-empty }\mathcal{K} \subseteq \{1,2,\dotsc,L\} \setminus \{i\}} \mu^*( a(\mathcal{K})) \label{eq:non-empty}\\
&=  H( W_{\{1,2,\dotsc, L\} \setminus \{i\}} | W_i).
\end{align}
\end{subequations}

\subsection{For $1 \leq |\mathcal{S}| \leq L-2$:}

Consider some $\mathcal{S} \subset \{1,2,\dotsc, L\}$ where $1 \leq |\mathcal{S}| \leq L-2$. We now show that if the ABCMI is satisfied, then
\begin{subequations}
\begin{align}
\sum_{i \in \mathcal{S}} r_i  &\geq H(W_\mathcal{S} | W_{\mathcal{S}^\text{c}}) \label{eq:second-case}\\
&= \sum_{\text{non-empty }\mathcal{K} \subseteq \mathcal{S}} \mu^*( a(\mathcal{K})).
\end{align}
\end{subequations}


Define $S=|\mathcal{S}|$ and $\boldsymbol{r}^\mathcal{S} \triangleq \{r_i : i \in \mathcal{S}\}$.
The LHS of \eqref{eq:second-case} is the sum of contributions from all atoms in $\mathbb{A}$ to all $r_i \in \boldsymbol{r}^\mathcal{S}$. We will divide all atoms in $\mathbb{A}$ according to their weight $K$: (i) $K=1$, (ii) $2 \leq K \leq S$, and (iii) $K \geq S+1$. So, we have 
\begin{align}
\sum_{i \in \mathcal{S}} r_i  &= \sum_{i \in \mathcal{S}} \sum_{\substack{\mathcal{K} \in \mathbb{K} \\ \text{s.t. } |\mathcal{K}| = 1}} J_i(\mathcal{K}) + \sum_{i \in \mathcal{S}} \sum_{K=2}^{S} \sum_{\substack{\mathcal{K} \in \mathbb{K} \\ \text{s.t. } |\mathcal{K}| = K}} J_i(\mathcal{K}) \nonumber \\ &\quad + \sum_{i \in \mathcal{S}} \sum_{K=S+1}^{L-1} \sum_{\substack{\mathcal{K} \in \mathbb{K} \\ \text{s.t. } |\mathcal{K}| = K}} J_i(\mathcal{K}). \label{eq:all-atoms}
\end{align}

\subsubsection{Atoms with weight $K=1$}

%

For atoms with weight one,
\begin{equation}
J_i (\mathcal{K}) = \begin{cases}
\mu^* ( a(\mathcal{K})), & \text{if } \mathcal{K} = \{i \} \\
0, & \text{otherwise.}
\end{cases} \label{eq:single}
\end{equation}
Summing the contributions from all atoms with weight one to all $r_i \in \boldsymbol{r}^\mathcal{S}$,
\begin{equation}
\sum_{i \in \mathcal{S}} \sum_{\substack{\mathcal{K} \in \mathbb{K} \\ \text{s.t. } |\mathcal{K}| = 1}} J_i(\mathcal{K}) = \sum_{\substack{\mathcal{K} \subseteq \mathcal{S} \\ \text{s.t. } |\mathcal{K}| = 1}} \mu^* ( a(\mathcal{K})). \label{eq:k-equals-1}
\end{equation}

%

\subsubsection{Atoms with weight $2 \leq K \leq S$}

We fix $K$. Consider the contributions from all atoms with weight $K$ to a particular $r_i$, $i \in \mathcal{S}$ (one column in the hashed region in Fig.~\ref{fig:atoms}). There are
\begin{itemize}
\item[{[}$\mathsf{O1}${]}] $\binom{L-1}{K-1}$ contributions with coefficient $\frac{L-K}{L-1}$ [from atoms where $i \in \mathcal{K}$; there are $\binom{L-1}{K-1}$  ways to select the other $(K-1)$ elements in $\mathcal{K}$ from $\{1,2,\dotsc, L\} \setminus \{i\}$], and
\item[{[}$\mathsf{O2}${]}] $\binom{L-1}{L-K-1}$ contributions with coefficient $-\frac{K-1}{L-1}$ [from atoms where $i \in \mathcal{K}^\text{c}$; there are $\binom{L-1}{L-K-1}$  ways to select the other $(L-K-1)$ elements in $\mathcal{K}^\text{c}$ from $\{1,2,\dotsc, L\} \setminus \{i\}$].
\end{itemize}
There are $\binom{L}{K}$ atoms with weight $K$. We can check that $\binom{L}{K} = \binom{L-1}{K-1} + \binom{L-1}{L-K-1}$.

Since observations $\mathsf{O1}$ and $\mathsf{O2}$ are true for each $r_i \in \boldsymbol{r}^{\mathcal{S}}$, we have the following contributions from all atoms with weight $K$ to $\boldsymbol{r}^\mathcal{S}$ (the hashed region in Fig.~\ref{fig:atoms}): There are
\begin{itemize}
\item[{[}$\mathsf{O3}${]}] $S\cdot\binom{L-1}{K-1}$ contributions with coefficient $\frac{L-K}{L-1}$, and
\item[{[}$\mathsf{O4}${]}] $S\cdot\binom{L-1}{L-K-1}$ contributions with coefficient $-\frac{K-1}{L-1}$.
\end{itemize}

For an atom $a(\mathcal{K})$, we say that the atom is {\em active} if $\mathcal{K} \subseteq \mathcal{S}$; otherwise, i.e., $\mathcal{K} \nsubseteq \mathcal{S}$, the atoms is said to be {\em inactive},

Consider the contributions from a particular active atom $a(\mathcal{K})$ to $\boldsymbol{r}^\mathcal{S}$ (one row in the hashed region in Fig.~\ref{fig:atoms}). We have the following contributions from this atom to $\boldsymbol{r}^\mathcal{S}$:
\begin{itemize}
\item[{[}$\mathsf{O5}${]}] Since $\mathcal{K} \subseteq \mathcal{S}$, $K$ cells in the hashed row each contribute $\frac{L-K}{L-1}\mu^*(a(\mathcal{K}))$.
\item[{[}$\mathsf{O6}${]}] The remaining $(S-K)$ cells each contribute $-\frac{K-1}{L-1}\mu^*(a(\mathcal{K}))$.
\end{itemize}
Summing the contributions from this active atom to $\boldsymbol{r}^\mathcal{S}$,
\begin{subequations}
\begin{align*}
\sum_{i \in \mathcal{S}} J_i(\mathcal{K}) &= \left[ K\frac{L-K}{L-1} - (S-K)\frac{K-1}{L-1} \right] \mu^*(a(\mathcal{K})) \nonumber \\
&= \left[ 1 + \frac{(L-S-1)(K-1)}{L-1} \right] \mu^*( a(\mathcal{K})).
\end{align*}
\end{subequations}

For any fixed $\mathcal{S}$, there are $\binom{S}{K}$ active atoms with weight $K$ (different ways of choosing $\mathcal{K} \subseteq \mathcal{S}$), and observations $\mathsf{O5}$ and $\mathsf{O6}$ are true for each active atom. Combining $\mathsf{O3}$--$\mathsf{O6}$, we can further categorize the contributions from all atoms with weight $K$ to $\boldsymbol{r}^\mathcal{S}$:
\begin{itemize}
\item[{[}$\mathsf{O7}${]}] Out of the $S\cdot\binom{L-1}{K-1}$ contributions with coefficient $\frac{L-K}{L-1}$, $K \cdot \binom{S}{K}$ of them are from active atoms.
\item[{[}$\mathsf{O8}${]}] Out of the $S\cdot\binom{L-1}{L-K-1}$ contributions with coefficient $-\frac{K-1}{L-1}$, $(S-K)\binom{S}{K}$ of them are from active atoms.
\end{itemize}

Now, summing the contributions from all (active and inactive) atoms with weight $K$ to $\boldsymbol{r}^\mathcal{S}$, we have
\begin{subequations}
\begin{align}
&\sum_{i \in \mathcal{S}} \sum_{\substack{\mathcal{K} \in \mathbb{K} \\ \text{s.t. } |\mathcal{K}| = K}} J_i(\mathcal{K}) = \sum_{i \in \mathcal{S}} \sum_{\substack{\mathcal{K} \subseteq \mathcal{S} \\ \text{s.t. } |\mathcal{K}| = K}} J_i(\mathcal{K}) + \sum_{i \in \mathcal{S}} \sum_{\substack{\mathcal{K} \nsubseteq \mathcal{S}\\ \text{s.t. } |\mathcal{K}| = K}} J_i(\mathcal{K}) \nonumber \\
&= \sum_{\substack{\mathcal{K} \subseteq \mathcal{S}\\ \text{s.t. } |\mathcal{K}| = K}} \left[ 1 + \frac{(L-S-1)(K-1)}{L-1} \right] \mu^*( a(\mathcal{K})) \nonumber \\ &\quad + \sum_{i \in \mathcal{S}} \sum_{\substack{\mathcal{K} \nsubseteq \mathcal{S}\\ \text{s.t. } |\mathcal{K}| = K}} J_i(\mathcal{K}) \nonumber \\
& \geq \sum_{\substack{\mathcal{K} \subseteq \mathcal{S}\\ \text{s.t. } |\mathcal{K}| = K}} \mu^*( a(\mathcal{K})) + \binom{S}{K} \frac{(L-S-1)(K-1)}{L-1} \underline{\mu}_K \nonumber \\
&\quad + \left[ S \cdot \binom{L-1}{K-1} - K \cdot \binom{S}{K} \right] \frac{L-K}{L-1} \underline{\mu}_K \nonumber \\
&\quad + \left[ S \cdot \binom{L-1}{L-K-1} - (S-K) \cdot \binom{S}{K} \right] \frac{-(K-1)}{L-1} \overline{\mu}_K \label{eq:eta} \\
& = \sum_{\substack{\mathcal{K} \subseteq \mathcal{S}\\ \text{s.t. } |\mathcal{K}| = K}} \mu^*( a(\mathcal{K})) + \frac{K-1}{L-1}\eta, \label{eq:k-greater-2}
\end{align}
\end{subequations}
where 
\begin{equation*}
\eta \triangleq \left[\alpha(L,S,K) + \frac{\beta(L,S,K)}{K-1} \right]\underline{\mu}_K  -\alpha(L,S,K) \overline{\mu}_K,
\end{equation*}
where $\alpha(L,S,K)$ is defined in \eqref{eq:alpha}  and $\beta(L,S,K)$ in \eqref{eq:beta}. If the ABCMI condition is satisfied, then $\eta \geq 0$.

\subsubsection{Atoms with weight $S+1 \leq  K \leq L-1$}
Consider the contributions from all atoms with weight $K$ to $r_i$, for some $i$. From observations $\mathsf{O1}$ and $\mathsf{O2}$, we know that there are
\begin{itemize}
\item $\binom{L-1}{K-1}$ contributions with coefficient $\frac{L-K}{L-1}$, and
\item $\binom{L-1}{L-K-1}$ contributions with coefficient $-\frac{K-1}{L-1}$.
\end{itemize}

Summing these contributions, we have for any $i$ that
\begin{align}
&\sum_{\substack{\mathcal{K} \in \mathbb{K}  \\ \text{s.t. } |\mathcal{K}| = K}} J_i(\mathcal{K}) \nonumber\\ & \geq \binom{L-1}{K-1} \frac{L-K}{L-1} \underline{\mu}_K  + \binom{L-1}{L-K-1}\frac{-(K-1)}{L-1} \overline{\mu}_K   \nonumber \\ 
& = \frac{ (L-2)!}{(L-K-1)!K!} \left[K \underline{\mu}_K - (K-1)\overline{\mu}_K \right] \geq 0. \label{eq:because-abcmi}
\end{align}
Since $\alpha(L,S,K) \geq \beta(L,S,K) \geq 0$, the ABCMI condition implies that $\overline{\mu}_K \leq \left[ 1 + \frac{1}{K-1} \right] \underline{\mu}_K= \frac{K}{K-1}\underline{\mu}_K$ for all $K \in \{2,3,\dotsc, L-1\}$. Hence, the inequality in \eqref{eq:because-abcmi} follows.

\subsubsection{Combining the contributions of  $\mathbb{A}$ to $\boldsymbol{r}^\mathcal{S}$}

Substituting \eqref{eq:k-equals-1}, \eqref{eq:k-greater-2}, and \eqref{eq:because-abcmi} into \eqref{eq:all-atoms}, if the sources have ABCMI, then
\begin{align*}
\sum_{i \in \mathcal{S}} r_i &\geq  \sum_{\substack{\mathcal{K} \subseteq \mathcal{S} \\  \text{s.t. } |\mathcal{K}| = 1}} \mu^* ( a(\mathcal{K}))  + \sum_{K=2}^{S}\, \sum_{\substack{\mathcal{K} \subseteq \mathcal{S} \\ \text{s.t. } |\mathcal{K}| = K}} \mu^* ( a(\mathcal{K})) \nonumber \\ & = H(W_\mathcal{S} | W_{\mathcal{S}^\text{c}}).
\end{align*}
This completes the proof of Theorem~\ref{theorem:tuple}. $\hfill \blacksquare$

\section{Proof of Theorem~\ref{theorem:main}} \label{sec:2}


\subsection{Necessary Conditions} \label{section:lower-bound}
\begin{lemma} \label{lemma:lower-bound}
Consider a finite field MWRC with correlated sources. A rate $\kappa$ is achievable only if \eqref{eq:main} holds for all $i \in \{1,2,\dotsc,L\}$.
\end{lemma}


Lemma 1 can be proved by generalizing the converse theorem~\cite[Appx.\ A]{ongtimo11submitted} for $L=3$ to arbitrary $L$. The details are omitted.

\subsection{Sufficient Conditions} \label{section:achievability}
Consider a separate source-channel coding architecture.

\subsubsection{Source Coding Region}

We have the following source coding result for correlated sources:
\begin{lemma} \label{lemma:source-coding}
Consider $L$ correlated sources as defined in Section~\ref{sec:sources}.
Each user $i$ encodes its message $\boldsymbol{W}_i$ to an index $M_i' \in \{1,2,\dotsc, 2^{m r_i}\}$, for all $i$. It reveals its index to all other users. Using these indices and its own message, each user $i$ can then decode the messages of all other users, i.e., $\{\boldsymbol{W}_j: \forall j \in \{1,2,\dotsc,L\} \setminus \{i\}\}$, if \eqref{eq:lemma} is satisfied
for all non-empty strict subsets $\mathcal{S} \subset \{1,2,\dotsc,L\}$. 
\end{lemma}

The above result is obtained by combining the results for (i) source coding for correlated sources~\cite[Thm.\ 2]{cover75}, and (ii) the three-user noiseless MWRC~\cite[Sec.\ II.B.1]{wynerwolf02}. Note that the relay does not participate in the source code, in contrast to the setup of~\cite{wynerwolf02}. Instead, the relay participates in the channel code. 


\subsubsection{Channel Coding Region}

We have the following channel coding result for the finite field MWRC~\cite{ongmjohnsonit11}: 
\begin{lemma}
Consider the finite field MWRC defined in \eqref{eq:uplink}--\eqref{eq:downlink}.
Let the message of each user $i$ be $M_i$, which is i.i.d. and uniformly distributed on $\{1,\dotsc, 2^{nR_i}\}$. Using $n$ uplink and downlink channel uses, each user $i$ can reliably decode the message of all other users $\{M_j: \forall j \in \{1,2,\dotsc, L\} \setminus \{i\} \}$ if
\begin{equation}
\sum_{ j \in \{1,2,\dotsc, L\} \setminus \{i\}} R_j \leq \log_2|\mathcal{F}| - \max\{ H(N_0), H(N_i)\}, \label{eq:channel-coding-rates}
\end{equation}
for all $i \in \{1,2,\dotsc, L\}$.
\end{lemma}

\subsubsection{Achievable Rates}

We propose the following separate source-channel coding scheme. Fix $m r_i = n R_i$.
Let $(D_1,\dotsc,D_L)$, where each $D_i$ is i.i.d. and uniformly distributed on $\{1,\dotsc,2^{mr_i}\}$. We call $D_i$ a \emph{dither}. The dithers are made known to all users. User $i$ performs source coding to compresses its message $\mathbf{W}_i$ to an index $M_i' \in \{1,\dotsc, 2^{m r_i}\}$ and computes $M_i = M_i' + D_i \mod 2^{mr_i}$. The random variables $(M_1,\ldots,M_L)$ are independent, with $M_i$ being uniformly distributed on $\{1,\dotsc,2^{n R_i}\}$.
The nodes then perform channel coding with $M_i$ as inputs. If \eqref{eq:channel-coding-rates} is satisfied, then each user $i$ can recover $\{M_j: \forall j \in\{1,\dotsc,L\} \setminus \{i\}\}$, from which it can obtain $\{M_j': \forall j \in\{1,\dotsc,L\} \setminus \{i\}\}$.  If  \eqref{eq:lemma}  is satisfied, then each user $i$ can recover all other users' messages $\{\boldsymbol{W}_j: \forall j \in \{1,\dotsc, L\} \setminus \{i\} \}$, using the indices $M_j'$'s  and its own message $\boldsymbol{W}_i$. This means the rate $\kappa = n/m$ is achievable.

Using the above coding scheme, we have the following:
\begin{lemma} \label{lemma:achievability}
Consider a finite field MWRC with correlated sources. If there exist a tuple $(r_1,r_2,\dotsc, r_L)$ and a positive real number $\kappa$ such that \eqref{eq:lemma} is satisfied for all non-empty strict subsets $\mathcal{S} \subset \{1,2,\dotsc, L\}$, and \eqref{eq:channel-coding-rates} is satisfied for all $i \in \{1,2,\dotsc, L\}$ with $R_i = r_i/ \kappa$, then the rate $\kappa$ is achievable.
\end{lemma}

\subsection{Proof of Theorem~\ref{theorem:main} (Necessary and Sufficient Conditions)}

The ``only if'' part follows directly from Lemma~\ref{lemma:lower-bound} (regardless of whether the sources have ABCMI). We now prove the ``if'' part. Suppose that the sources have ABCMI. From Theorem~\ref{theorem:tuple}, there exists a tuple $(r_1,\dotsc, r_L)$ such that conditions $\mathsf{C1}$ and $\mathsf{C2}$ are satisfied. 
These conditions imply that \eqref{eq:lemma} is satisfied for all non-empty strict subsets $\mathcal{S} \subset \{1,2,\dotsc, L\}$. Let $R_i = r_i/ \kappa$. Condition $\mathsf{C2}$ further implies $\kappa = \frac{H(W_{\{1,2,\dotsc,L\}}|W_i)}{\sum_{i \in \{1,2,\dotsc, L\} \setminus \{i\}} R_j}$. So, if \eqref{eq:main} is true, then \eqref{eq:channel-coding-rates} is satisfied for all $i \in \{1,2,\dotsc,L\}$. From Lemma~\ref{lemma:achievability},  $\kappa$ is achievable. $\hfill \blacksquare$.

\begin{remark}
For a fixed source correlation structure and a finite field MWRC, one can check if the $L$-dimensional polytope defined by the source coding region and that by the channel coding region (scaled by $\kappa$) intersect when $\kappa$ equals the RHS of \eqref{eq:main}. If the regions intersect, then we have the set of all achievable $\kappa$ for this particular  source-channel combination. Theorem~\ref{theorem:main} characterizes a class of such sources.
\end{remark}



\end{document}